\def\papertitle{Synthesizer Sound Matching Using Audio Spectrogram Transformers} 
\def\paperauthorA{Fred Bruford}
\def\paperauthorB{Frederik Blang}
\def\paperauthorC{Shahan Nercessian}

\documentclass[twoside,a4paper]{article}
\usepackage{etoolbox}
\usepackage{dafx_24_lbr}
\usepackage{amsmath,amssymb,amsfonts,amsthm}
\usepackage{euscript}
\usepackage[T1]{fontenc}
\usepackage[utf8]{inputenc}
\usepackage{ifpdf}
\usepackage[english]{babel}
\usepackage{caption}
\usepackage{subfig} 
\usepackage{color}

\input glyphtounicode
\ninept

\newcounter{numauth}
\newcounter{listcnt}
\newcommand\authcnt[1]{\ifdefined#1 \stepcounter{numauth} \fi}

\newcommand\addauth[1]{
\ifdefined#1 
\stepcounter{listcnt}
\ifnum \value{listcnt}<\value{numauth}
\appto\authorslist{, #1}
\else
\appto\authorslist{~and~#1}
\fi
\fi}
\authcnt{\paperauthorB}
\authcnt{\paperauthorC}
\authcnt{\paperauthorD}
\authcnt{\paperauthorE}
\authcnt{\paperauthorF}
\authcnt{\paperauthorG}
\authcnt{\paperauthorH}
\authcnt{\paperauthorI}
\authcnt{\paperauthorJ}
\def\authorslist{\paperauthorA}
\addauth{\paperauthorB}
\addauth{\paperauthorC}
\addauth{\paperauthorD}
\addauth{\paperauthorE}
\addauth{\paperauthorF}
\addauth{\paperauthorG}
\addauth{\paperauthorH}
\addauth{\paperauthorI}
\addauth{\paperauthorJ}

\usepackage{times}

\newif\ifpdf
\ifx\pdfoutput\relax
\else
   \ifcase\pdfoutput
      \pdffalse
   \else
      \pdftrue
   \fi
\fi

\ifpdf 
  \usepackage[pdftex,
    pdftitle={\papertitle},
    pdfauthor={\authorslist},
    pdfsubject={Proceedings of the 27th International Conference on Digital Audio Effects (DAFx24)},
    colorlinks=false, 
    bookmarksnumbered, 
    pdfstartview=XYZ 
  ]{hyperref}
  \usepackage[pdftex]{graphicx}
\else 
  \usepackage[dvips]{epsfig,graphicx}
  \usepackage[dvips,
    pdftitle={\papertitle},
    pdfauthor={\authorslist},
    pdfsubject={Proceedings of the 27th International Conference on Digital Audio Effects (DAFx24)},
    colorlinks=false, 
    bookmarksnumbered, 
    pdfstartview=XYZ 
  ]{hyperref}
\fi
\usepackage[hypcap=true]{caption}
\title{\papertitle}

\affiliation{
\paperauthorA\,\sthanks{Thanks to the predecessors for the templates}}
{\href{https://dafx24.surrey.ac.uk}{Institute of Sound Recording} \\ University of Surrey\\ Guildford, UK\\
{\tt \href{mailto:dafx24@surrey.ac.uk}{dafx24@surrey.ac.uk}}
}
\affiliation{
\paperauthorA$^{1}$, \paperauthorB$^{2}$, and \paperauthorC$^{3}$\thanks{\vspace{-3mm}}}
{\href{https://www.native-instruments.com/}{Native Instruments} \\ $^{1}$London, United Kingdom $^{2}$Berlin, Germany $^{3}$Boston, MA, USA \\
{\tt \{\href{mailto:fred.bruford@native-instruments.com}{fred.bruford}|\href{mailto:frederik.blang@native-instruments.com}{frederik.blang}|\href{mailto:shahan.nercessian@native-instruments.com}{shahan.nercessian}\}@native-instruments.com}
}

\begin{document}
\ifpdf 
  \DeclareGraphicsExtensions{.png,.jpg,.pdf}
\else  
  \DeclareGraphicsExtensions{.eps}
\fi


\maketitle

\begin{abstract}

Systems for synthesizer sound matching, which automatically set the parameters of a synthesizer to emulate an input sound, have the potential to make the process of synthesizer programming faster and easier for novice and experienced musicians alike, whilst also affording new means of interaction with synthesizers. Considering the enormous variety of synthesizers in the marketplace, and the complexity of many of them, general-purpose sound matching systems that function with minimal knowledge or prior assumptions about the underlying synthesis architecture are particularly desirable. With this in mind, we introduce a synthesizer sound matching model based on the Audio Spectrogram Transformer. We demonstrate the viability of this model by training on a large synthetic dataset of randomly generated samples from the popular \textit{Massive} synthesizer. We show that this model can reconstruct parameters of samples generated from a set of 16 parameters, highlighting its improved fidelity relative to multi-layer perceptron and convolutional neural network baselines. We also provide audio examples demonstrating the out-of-domain model performance in emulating vocal imitations, and sounds from other synthesizers and musical instruments.

\end{abstract}

\section{Introduction and Related Work}

Systems for controlling synthesizers based on an input sound, often referred to as \textit{synthesizer sound matching}, \textit{parameter estimation} or \textit{automatic synthesizer programming}, have attracted research interest since at least the 1990s \cite{horner1993machine, wehn1998using} due to their practical music production applications. Tools that manipulate the parameters of a synthesizer automatically to recreate an input sound could make programming synthesizers more accessible for musicians with a limited understanding of sound synthesis, but also speed up synthesizer programming for experienced musicians alike. Beyond its use as a workflow improvement, synthesizer sound matching could open up new creative possibilities, enabling the use of sound as a control interface to synthesizers, and aiding the creation of new and unique synthesizer patches beyond factory preset libraries. As some rudimentary examples, synthesizer sound matching could open up the possibility of controlling synthesizers through query-by-vocalization, or enable users to recreate sounds from sampled tracks using their own synthesizer.

Systems for synthesizer sound matching fall on a scale according to the level of knowledge they assume of the underlying synthesis architecture. Many contemporary solutions \cite{masuda,ddx7, diffmoog} require a fully differentiable implementation of the synthesizer to be a part of the model, assuming that leveraging knowledge of the sound synthesis process should make synthesizer sound matching easier. However, such architectures are inflexible in that they will only work for the synthesizers that they are explicitly designed for, and new architectures will have to be developed for each individual synthesizer. This also becomes intractable as the complexity of the synthesizers being modelled increases; it may be possible to implement simpler synthesizers with differentiable operations, but to do so for the highly complex synthesizers that find widespread use in contemporary music production practice, utilizing banks of wavetables, effects, and complex modulation and routing options, could be prohibitively labor-intensive.  

As a result, many approaches to synthesizer sound matching attempt to bypass the use of differentiable processors altogether to create more scalable solutions for the problem. \textit{InverSynth} \cite{inversynth} considers convolutional neural network (CNN) architectures inferring synthesizer configurations from either input spectrograms or even raw audio, whereby models optimize an objective function defined over inferred and ground truth synthesizer parameters.  Acknowledging that parameter losses do not directly correlate to perception, \textit{FlowSynth} leverages a variational auto-encoder with normalizing flows in order to jointly organize and map auditory and parameter spaces \cite{flowsynth}.  Extensions have also considered the joint modeling of several virtual instruments concurrently, encouraging formulation of the method as a multi-task learning problem to further promote system generalization \cite{faronbi}.

\begin{figure*}[t]
    \centering{\includegraphics[width=0.95\textwidth]{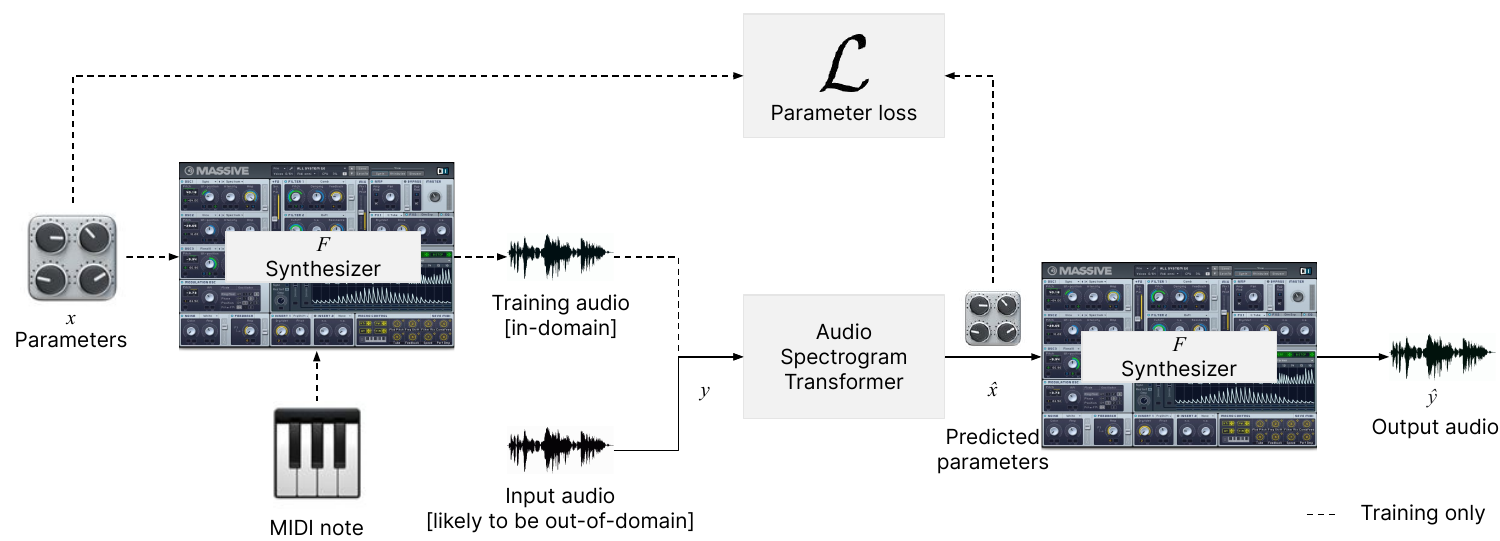}}
    \caption{Proposed synthesizer sound matching system block diagram.}
    \vspace{-1.0em}
    \label{fig:system}
\end{figure*}

In recent years, Transformers \cite{attention} have been finding increased usage in classification and regression tasks, beyond their popular use as generative models. This has led to their increased uptake within the domain of music information retrieval (MIR) specifically; recently architectures leveraging transformers have been found to outperform more traditional convolutional or recurrent architectures in archetypal MIR tasks such as music tagging \cite{ast, patchout} and piano music transcription \cite{piano}. However, this comes at the cost of increased computational complexity and data requirements, something that can be a problem for music-focused tasks where labeled data can be scarce.

In this paper, we propose a synthesizer sound matching method using an Audio Spectrogram Transformer (AST) \cite{ast} backbone.  To the best of our understanding, this is the first work to consider the use of an AST for the synthesizer sound matching problem. In doing so, we advance work towards a general-purpose synthesizer sound matching architecture that requires minimal assumptions about the underlying synthesis architecture. ASTs offer a natural solution to this problem due to their strong performance across supervised learning tasks in the music domain. Furthermore, through generation of randomly sampled datasets of paired parameter values and respective audio samples rendered through a synthesizer, we can leverage arbitrarily large, fully-labeled datasets that afford the successful training of transformers, a rarity in the context of MIR research. For demonstrative purposes, we evaluate our proposed model on \textit{Massive} by Native Instruments, a popular and widely used synthesizer, known for its expressivity and complexity. We carry out an automated evaluation of our synthesizer sound matching model for parameter prediction quality and audio reconstruction quality, comparing our approach against both multi-layer perceptron (MLP) and CNN baselines.  Lastly, we provide examples of sound matching model performance for both in- and out-of-domain audio inputs, available at \href{https://synth-ast-dafx.netlify.app}{\nolinkurl{https://synth-ast-dafx.netlify.app}}.

\section{Proposed Method}

Figure \ref{fig:system} outlines our proposed approach. We consider the problem of synthesizer sound matching as one of \textit{parameter inference}, whereby a predictive model (such as an AST), taking an audio sample $y$ as an input, attempts to predict the underlying synthesis parameters $x$ that were used to render that sample. These predicted parameters $\hat{x}$ can then be used to render a new sound $\hat{y}=F(\hat{x})$ that matches the input audio sample by means of a given synthesizer $F$. This inference process is shown on the right side of Figure \ref{fig:system}. To achieve this, a typical approach is to train deep learning models on paired samples of parameter sets $x$ and rendered one-shot samples $y = F(x)$. The training process is shown in the left side of Figure \ref{fig:system}. While such a model will be able to recreate synthesizer sounds from the synthesizer being modeled by design, we rely on these models having strong \textit{out-of-domain} performance to be able to generate parameters approximating an arbitrary audio example $y$ originating from any possible sound source.  This is to say that models should be robust to the fact that input audio at inference time is unlikely to be constrained to the subspace of sounds created via the signal model $y = F(x)$. 

\subsection{The Audio Spectrogram Transformer}

Our proposed model is derived from the Audio Spectrogram Transformer (AST) \cite{ast}, itself based primarily on the Vision Transformer (ViT) model \cite{vit}. AST is an architecture for audio classification based fundamentally on a vanilla Transformer encoder, combined with a patch-based Mel spectrogram pre-processor. 

Transformers typically take as input a 1D vector of token embeddings. To convert a single 2D Mel spectrogram to a 1D sequence of token embeddings, AST splits the spectrogram into a series of \textit{patches} across both axes of the spectrogram in a manner similar to that used in ViT for image data. We use \(16 \times 16\) patches with a stride of 6. Each patch is flattened to pass into a linear embedding layer of size \(768\). Due to their use of residual connections between layers, transformers require that all its internal layers are of the same size, so the internal dimensionality of AST is also \(768\). 

As in ViT, AST uses a learnt positional embedding which is summed with the patch embedding to retain the 2D position of each patch, before passing through to an encoder-only Transformer. It uses 12 encoder layers, with 12 attention heads each, and a hidden size of \(768\), copying the hyperparameters used in the smallest ViT model, ViT-Base \cite{vit}.

\subsection{AST for Synthesizer Sound Matching}

We make a number of modifications to AST to make it suitable for the problem of synthesizer sound matching. Firstly, to enable it to work for our regression formulation rather than classification, the final output layer is modified to use an mean-squared error (MSE) loss between output parameter values and target parameter values, rather than the binary cross-entropy loss used in AST. As an additional improvement, we insert a small 3-layer MLP of width 768 between the output of the Transformer encoder and the output layer. We found this to increase the performance of the model, with a minimal effect on training time.

Secondly, AST applies an adaptation to the learnt positional encoding that allow it to model sequences of variable length by interpolating between embeddings. Allowing for variable length inputs is less important for prototyping the synthesizer sound matching, as we assume all inputs sounds are one-shot samples, meaning we can neglect this adaptation and treat the input spectrograms as fixed-size, the same as images are treated in ViT.

We also reduce the size of our Mel spectrogram to 64 bins rather than 128. We found this to have minimal impact on downstream performance, while halving the number of patches required to capture a given spectrogram, thus significantly speeding up the training time. Finally, both ViT and AST pre-train on ImageNet \cite{deng2009imagenet} before fine-tuning on downstream classification tasks, but due to our ability to create datasets of arbitrary size (as discussed in Section \ref{sec:dataset}) we do not carry this out.

\section{Experimental Results}
\subsection{Dataset}
\label{sec:dataset}

Our dataset for the synthesizer sound matching task consists of paired examples of parameter settings and the respective rendered one-shot samples for our chosen synthesizer, \textit{Massive}. Rather than deriving samples from factory presets (as in \cite{flowsynth} or \cite{faronbi}), we create a synthetic dataset by randomly sampling parameter values and generating corresponding one-shots, an approach also used in \cite{inversynth}. This allows us to sample the parameter space of \textit{Massive} at arbitrary levels of complexity by varying only desired parameters, and generate datasets of an arbitrary size, allowing for the scale required for training the Transformer architecture.

We automate the generation of this dataset using the \textit{Pedalboard} Python library \cite{sobot_peter_2023_7817838}, feeding \textit{Massive} with random parameter values and rendering the output audio for a note of pitch C4 and maximum velocity. Each generated one-shot sample is 4 seconds long, consisting of a 1-second MIDI note on event at the beginning of the sample. Any samples generated with a loudness below a threshold of -60dB were removed to ensure there were no silent or near-silent samples. One disadvantage of using random datasets generated in this way is that the resulting samples may be less practical or realistic than typical synthesizer sounds that a user may want to create. To alleviate this, rather than sampling parameter values from a uniform distribution, we sample them from within the distribution of parameters extracted across all the \textit{Massive} factory presets, therefore sampling more of the `usable' subspace of parameter values that is found in practice.

We created a dataset of 1 million paired samples using a set of 16 parameters. The parameters were selected by hand to lead to sonically diverse samples from a scaled-back parameter set in order to provide an easier proof-of-concept. Random samples are created based on a `base' preset using two oscillators (one with fixed pitch), two filters, two envelopes and a reverb. We select only continuous parameters at this stage to simplify the required architecture, avoiding any categorical parameters or binary switches. The parameters used are shown in Table \ref{tab:parameters}.

\begin{table}[t]
  \caption{\itshape Massive parameters used for 16 parameter dataset}
	\centering
	\begin{tabular}{|c|l|}
    	\hline
     Parameter name & Description \\
     \hline
            osc1\_position  & Wavetable start position for oscillator 1 \\
           osc1\_amp  & Output gain of oscillator 1 \\
           osc2\_pitch & Detune pitch of oscillator 2\\
            osc2\_position & Wavetable start position for oscillator 2 \\
           osc2\_amp  & Output gain of oscillator 2 \\
            noise\_amp & Noise generator output gain \\
          filter1\_cut\_prm\_1 & Lowpass filter cutoff\\
          filter1\_res\_prm\_3 & Lowpass filter resonance\\
          filter2\_cut\_prm\_1 & Highpass filter cutoff\\
          filter2\_res\_prm\_3 & Highpass filter resonance\\
          envelope4\_att\_tme & Output amplitude attack\\
          envelope4\_dec\_tme & Output amplitude decay\\
          envelope1\_att\_tme & Lowpass filter cutoff attack\\
          envelope1\_att\_lev & Lowpass filter cutoff envelope level\\
          envelope1\_dec\_tme & Lowpass filter cutoff decay \\
            master\_fx2\_dry\_wet & Reverb dry/wet \\\hline
             
	\end{tabular}
	\label{tab:parameters}
 \vspace{-1.0em}
\end{table}

We extract and store 64-bin Mel spectrograms from each rendered sample, resulting in a total dataset size of 22GB. Running multiple instances of \textit{Massive} in parallel via \textit{Pedalboard}, we are able to generate this dataset in approximately 24 hours on a dedicated \textit{Mac mini} (2018).

\subsection{Evaluation Method}
As in \cite{flowsynth}, we evaluate our models on the test set for both parameter reconstruction and audio reconstruction accuracy using two respective measures: mean-squared error (MSE) between predicted and input parameters, and \textit{Spectral Convergence} (SC) \cite{spectral} between input audio and output audio as rendered through \textit{Massive} from the input and predicted parameter sets. The latter allows us to directly measure the audio quality of the synthesizer sound matching, which may not be reflected by the parameter reconstruction score alone. In order to further underscore the effectiveness of our methods, we compare our AST model against two established baselines. All SC and MSE scores are aggregated over 10,000 examples taken from the test set.

Aside from quantitative evaluation metrics, to subjectively evaluate both in-domain and out-of-domain sound matching model performance we provide audio examples of reconstructed sounds at our accompanying website. In-domain samples are \textit{Massive} presets drawn from the test set, while out-of-domain samples include vocal synthesizer imitations, sounds from other synthesizers, and musical instrument one-shots. As the AST outputs a set of parameters which can be rendered at any MIDI pitch, reconstructed sounds are rendered at the pitch deduced from input audio.

\subsubsection{Baselines}
Our two baseline models are implemented based on those used in \cite{flowsynth}, and are each trained on the same dataset for 50 epochs.
\begin{itemize}
    \item[\textbf{MLP}] 5-layer MLP with width of 2048, including batch normalization and a dropout of \(p=0.3\), with ReLU activations.
    \item[\textbf{CNN}] 5-layer, 128-channel 2D CNN with kernel size \(7\), stride \(2\), batch normalization and dropout of \(p=0.3\). This is followed by a 5-layer MLP using the same hyperparameters as the standalone MLP.
\end{itemize}

Our 1 million sample dataset was split into training, validation and test sets in a 80/10/10 split. All models were trained for 50 epochs using the Adam optimizer, though the CNN and MLP training converged after roughly 5 epochs. The learning rate of the CNN and MLP training was \( 3e-4\) and the AST \(5e-5\). Evaluation was carried out on checkpoints with minimum validation loss.

\subsection{Results}
As can be seen in Table \ref{tab:results}, our AST model significantly outperforms both baselines in both parameter prediction accuracy (MSE) and audio reconstruction accuracy (SC), demonstrating its strength at predicting parameters for in-domain data, with a relatively small difference between our two baselines. The strong performance of the AST over the baselines in terms of audio reconstruction accuracy shows that the ability to reconstruct parameter sets does result in well reconstructed one-shots.

\begin{table}[htb]
  \vspace{-0.0em}
  \caption{\itshape Results for parameter reconstruction mean-squared error (\textit{MSE}) and audio reconstruction spectral convergence (\textit{SC}).}
	\centering
	\begin{tabular}{|c|c|c|c|}
    	\hline
              & \textbf{MLP} & \textbf{CNN} & \textbf{AST} \\
            \hline
            MSE & 0.077 & 0.094 & \textbf{0.031} \\
            SC & 4.608 & 5.372 & \textbf{0.616} \\\hline
	\end{tabular}
	\label{tab:results}
  \vspace{-1.0em}
\end{table}

Subjectively, the audio demos also reflect the strong performance of the AST-based sound matching model. For in-domain sounds, the AST model is broadly able to reconstruct the timbre and envelope of the input sound, albeit limited in expressivity and accuracy by the constrained set of 16 parameters it operates on. The AST is also able to reconstruct timbre and envelope for out-of-domain input sounds, suggesting this approach can be used to approximate arbitrary audio examples effectively, despite using only training data extracted from the synthesizer. One failure mode of the model seems to be in the modelling of oscillator pitch. This effect can be heard in Sample 8 and Sample 11 of the in-domain examples. In both of these, one oscillator is detuned, while the rest of the sample is reconstructed accurately. Oscillator pitch is one parameter where the discrepancy between parameter difference and sonic difference can be very high; a small difference in the tuning of one oscillator can introduce beating or dissonance, significantly affecting how a sample sounds. These kinds of issues could be a limitation of using a parameter-only loss function rather than incorporating an audio-based loss function, or a limitation of our data rendering and training strategy, where we do not vary the MIDI pitch of sounds in our training dataset, and do not train our model explicitly to predict the overall pitch of an input sound alongside other parameters.

\section{Conclusion}
In this paper, we provide some early evidence that our AST-based architecture, trained on a large synthetic dataset of randomly sampled one-shots from a software-based synthesizer, can yield strong performance in a synthesizer sound matching task. This could indicate the viability of such an architecture for general-purpose synthesizer sound matching, where sound matching systems could be created for further synthesizers with minimal architectural modifications, and without requiring them to be implemented differentiably. The ability of our AST-based architecture to generalize effectively also suggests arbitrary input sounds can be approximated effectively even when the training data is derived solely from the synthesizer itself.

As further work, we hope to scale our models to larger and more complex sets of synthesizer parameters and audio data. For more comprehensive modeling of synthesizers, we aim to investigate multi-output models that concurrently model continuous and categorical parameters, thus capturing the different kinds of parameters typically seen in synthesizers. We also aim to investigate specific architectural modifications for modeling routing and modulation in the context of synthesizer parameter matching. Finally, we intend to experiment further with our dataset rendering strategy to enable our models to match the pitch of a sound more effectively, including rendering samples at multiple pitches, and learning input pitch explicitly as a parameter.

\section{Acknowledgments}
Thanks to the entire Audio Research team at Native Instruments for their support and feedback throughout the duration of this work. In particular, thanks to Johannes Imort for his assistance in setting up the demo website and Max Reifsteck for his assistance in dataset generation. Finally, thanks to the anonymous reviewers for their feedback on this work.

\nocite{*}
\bibliographystyle{IEEEbib}
\bibliography{DAFx24_tmpl} 
\end{document}